\documentclass{article}

\usepackage{arxiv}

\usepackage[utf8]{inputenc} % allow utf-8 input
\usepackage[T1]{fontenc}    % use 8-bit T1 fonts
\usepackage{hyperref}       % hyperlinks
\usepackage{url}            % simple URL typesetting
\usepackage{booktabs}       % professional-quality tables
\usepackage{amsfonts}       % blackboard math symbols
\usepackage{nicefrac}       % compact symbols for 1/2, etc.
\usepackage{microtype}      % microtypography
\usepackage{lipsum}
\usepackage[ruled,vlined]{algorithm2e}
\usepackage{graphicx}
\usepackage{cite}
\usepackage{subcaption}
\usepackage[square,numbers]{natbib}
\usepackage{setspace}

\title{Super-Resolution Remote Imaging using Time Encoded Remote Apertures}

\author{
  Ji Hyun Nam\\
  Department of Electrical and Computer Engineering\\
  University of Wisconsin Madison\\
  Madison, WI 53706 \\
  \texttt{namjey@gmail.com} \\
   \And
  Andreas Velten\\
  Department of Biostatistics and Medical Informatics\\
  University of Wisconsin Madison\\
  Madison, WI 53706 \\
  \texttt{velten@gmail.com} \\
}
\begin{document}
\maketitle

\begin{abstract}
Imaging of scenes using light or other wave phenomena is subject to the diffraction limit. The spatial profile of a wave propagating between a scene and the imaging system is distorted by diffraction resulting in a loss of resolution that is proportional with traveled distance. We show here that it is possible to reconstruct sparse scenes from the temporal profile of the wave-front using only one spatial pixel or a spatial average. The temporal profile of the wave is not affected by diffraction yielding an imaging method that can in theory achieve wavelength scale resolution independent of distance from the scene.

\end{abstract}

% keywords can be removed
\section{Introduction}
\subsection{Previous attempts to beat the diffraction limit of resolution}
The diffraction limit naturally limits the spatial resolution of an image acquired by a conventional imaging system with a finite aperture. A wave traveling from the scene to the imaging system is subject to diffraction. The wave reaching the aperture of the imaging system has been degraded by diffraction. It can no longer be used to reconstruct an image of the object at the optimal resolution. Any spatial wavefront is subject to this diffraction that causes the well known linear drop in resolution with distance. The resulting linear dependence of resolution on imaging distance is known as the Rayleigh criterion.%~\cite{rayleigh}

Over the last several years, many works in different fields have been done to beat the diffraction limit. Astronomical imaging is  area where super-resolution methods are extensively studied. Most of the super-resolution methods used in astronomy combine image capturing techniques with image post-processing. In early years co-addition methods, capturing successively multiple pictures within a short exposure time and applying shift-and-add algorithm~\cite{christou1991infrared} or linear reconstruction scheme~\cite{fruchter2002drizzle} to combine the images, were actively used. Different regularization-based optimization methods were developed: Orieux et al.\cite{orieux2012super} applied successfully applied quadratic-regularized optimization to the "Spectral and Photometric Imaging Receiver" data. Jarret et al\cite{jarrett2012extending} proposed a method using a deconvolution technique, maximum correlation method, combined with a re-sampling kernel to super-resolve "Wide-field Infrared Survey Explorer" data.  

Several image post processing methods exists in digital imaging applications to construct high-resolution images from low-resolution images. Nonuniform interpolation approach ~\cite{clark1985transformation}, frequency domain approach~\cite{tsai1989multiple}, regularized optimizations (deterministic~\cite{hong1997regularized,hong1997iterative,hardie1998high,bose2001advances} and stochastic~\cite{tom1995reconstruction,schultz1996extraction,hardie1997joint,cheeseman1996super}), iterative back-projection~\cite{irani1991improving}, adaptive filtering approach~\cite{elad1999superresolution} are widely used super-resolution imaging methods. These super-resolution methods provide high-resolution image outputs, mitigating the effects of diffraction. However, all method s still suffer from the fundamental limit: The resolution of super-resolved images still depends on the distance and aperture of the imaging system. The further the region of interest, the lower the image resolutions. 

It is well known that a scattering medium around an object can be used to couple evanescent waves into the far field. This means that images of the embedded object with resolutions similar to the wavelength can be captured from large distances ~\cite{ourir2012far}. In existing approaches, this requires prior characterization of the scattering medium from the location of the object, making the method unsuitable for many practical applications. Here we show that in sparse scenes, scattering can be leveraged to determine the geometry of the entire scattering scene (object and medium) without prior information based on just a single pixel time response of the scene.

\subsection{Our contribution}
In this work, we propose a novel super-resolution imaging approach that reconstructs scenes from internally scattered light using their time response. This temporal signal is not subject to diffraction. We show that the resolution of our method is independent of distance and aperture size, and we can reconstruct the scene up to Euclidean congruence. We demonstrate the capabilities of the novel approach using simulated and experimental data. 

\begin{figure}
  \includegraphics[width=\textwidth,height=8cm]{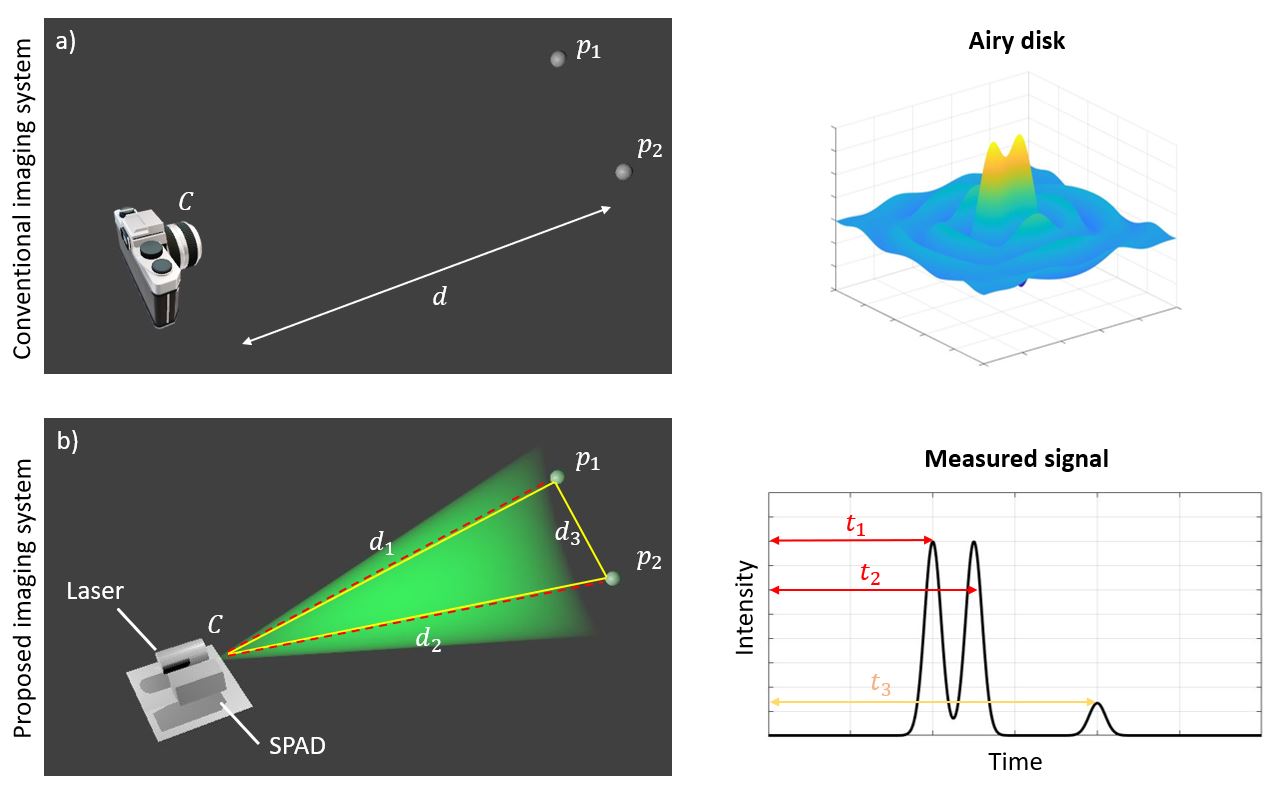}
  \caption{\textbf{Comparison between conventional and TERA imaging system.}
  \label{fig_comparison}
  \textbf{a)} Conventional imaging system. To infer the position of an object point within the focal plane, the imaging system needs to determine the difference in length between two rays. The figure shows a fundamental relation between distance, resolution, and aperture. The resolution of the system is determined by the Rayleigh's criterion.  
  \textbf{b)} TERA imaging system. A pulsed laser illuminates the scene; the light reflected back from the scene is detected by the SPAD. The collected time response contains information about the distance between the two targets. The resolution of the system is determined by the time resolution of the SPAD.  
	}
\end{figure}

\section{Time Encoded Remote Apertures (TERA)}
\subsection{Fundamentals of imaging}
In this section, we overview the fundamental process of obtaining an image from an object and define terms to avoid confusion. Consider an object emitting or reflecting light and an imaging system with an aperture size $d$ placed at a distance L from for the object (Figure \ref{fig_comparison}). To infer the position of an object point $O_i$ within the scene, the imaging system redirects all rays from point $O_i$ such that they constructively interfere on exactly one point in the focal plane. In other words, the imaging system evaluates the length of a light ray as encoded in its phase. In this case, the resolution $r$ of the imaging system is determined by the Rayleigh's diffraction limit:
    \begin{equation}
    r=\frac{1.22\lambda d}{D}
    \end{equation}

where $\lambda$ is the wavelength of the emitted light and determines how accurately the imaging system can determine the length of each ray, $d$ is a distance between focal plane of the imaging system and the object, and $D$ is the diameter of the aperture of the imaging system. 
Another way of imaging is to measure the time of flight of intensity fluctuation. We call these intensity fluctuations as second-order coherence.  One can use a short-pulsed laser to illuminate the object and lens-less time-of-flight detector to observe back-scattered light and reconstruct an image. The resolution  of such an image is described by the Rayleigh criterion, except that the wavelength $\lambda$ is replaced by the time resolution $\tau$ of the time-of-flight detector. We call it a transient Rayleigh criterion.

\begin{figure}
  \includegraphics[width=\linewidth,height=9cm]{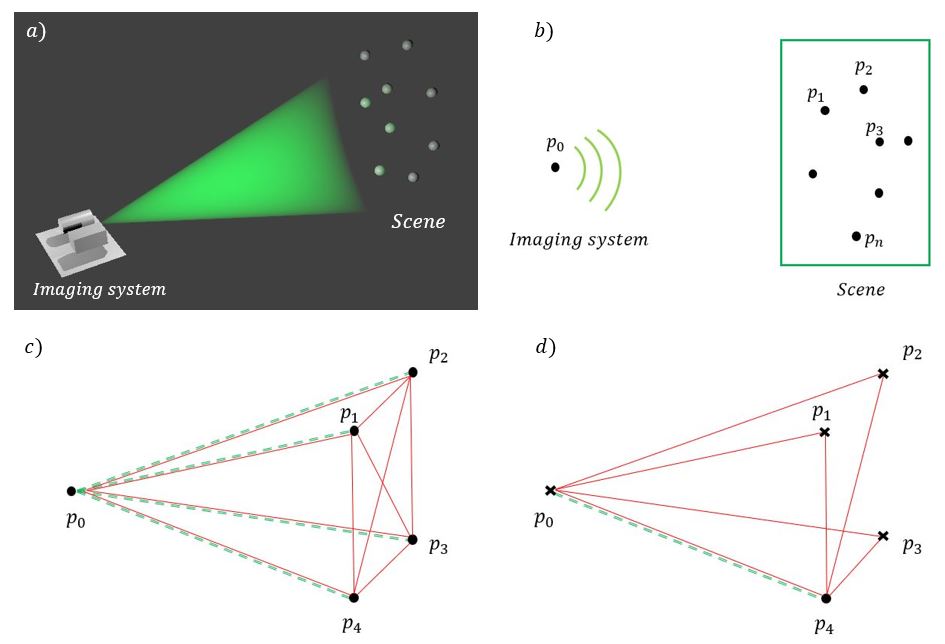}
  \caption{
  \label{bounce_path}
  \textbf{a)} Illustration of the imaging system(a pulsed laser and time-of-flight camera) and the point-cloud scene
  \textbf{b)} We consider the imaging system as part of the point cloud. $p_0$ is the location of the imaging system. $p_1, p_2, ..., p_n$ are point objects in the scene.
  \textbf{c)} Dashed green lines are pings, and solid red lines are loop paths. If the measurement ensemble $\beta$ contains four pings and six loops then we say that the sub-graph is contained in the measurement ensemble
  \textbf{d)} If "x" are known points and the measurement ensemble $\beta$ contains one ping and three loops, then we say that the measurement allows for trilateration.
	}
\end{figure}

\begin{figure}[!b]
  \includegraphics[width=\textwidth,height=7cm]{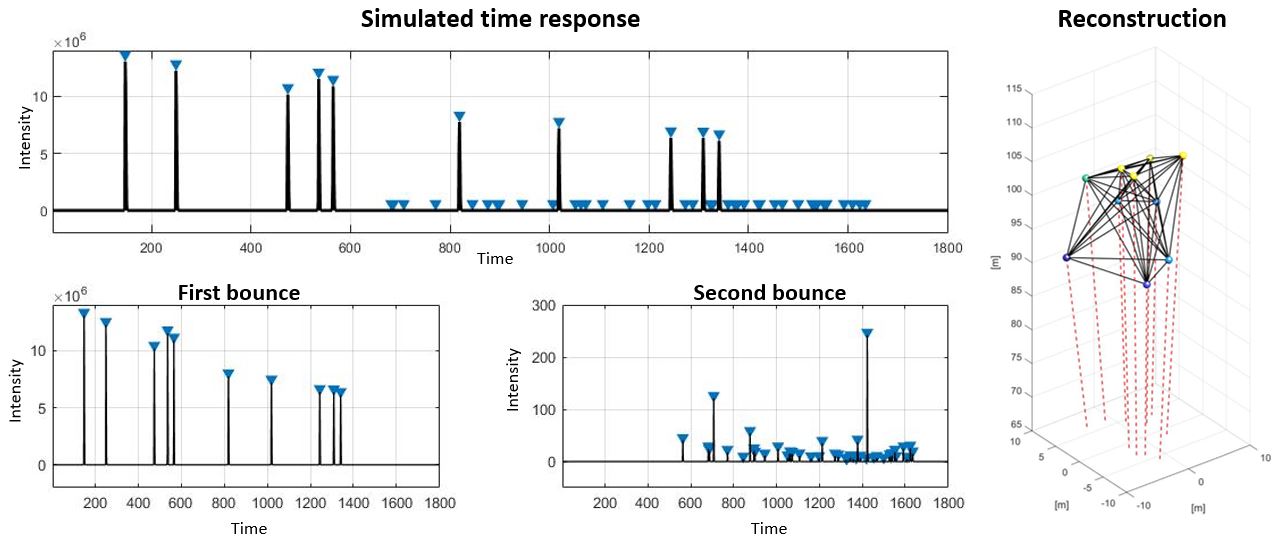}
  \caption{\textbf{Time response and reconstruction}
  \label{fig_timeresponse_reconstruction}
  \textbf{a)} Total time response of randomly generated scene with 10 points. The time response includes the first and second bounces. The peak extraction algorithm is applied to find peaks and marked in blue.
  \textbf{b)} Second bounce time response only;
  \textbf{c)} Reconstruction of the point-cloud. After applying a peak extraction algorithm, modified TRIBOND is used to reconstruct the point-cloud. The red dotted line corresponds to the first bounce path connecting point in the scene and imaging system. The black line is the second bounce path.
	}
\end{figure}

\subsection{TERA imaging overview}
Consider a situation when one can detect not only direct(first bounce) back-scattered lights but also the lights that bounced between the objects(second bounce). Such signal can be obtained by illuminating the scene with a short pulsed laser and detecting the returning light with a single time-of-flight detector. The temporal signal of these multi-bounce lights directly encodes the information about the distance between two object points in the scene. Figure \ref{fig_comparison}b) shows the simplified example of how the information about the scene is encoded in the temporal signal. Suppose the scene contains two small point objects $p_1$ and $p_2$ with negligibly small diameter $\delta$. The distance from $p_1$ and $p_2$ to the imaging system $C$ are $d_1$ and $d_2$ correspondingly. The distance between objects $p_1$ and $p_2$ is $d_3$. The scene is being illuminated by a pulsed laser and the returning light from the scene is measured by a time-of-flight detector. The time-resolved measurement or time response of the scene is illustrated on the left side of Figure \ref{fig_comparison}b). The first impulse is due to the light directly reflected from object $p_1$ and it appears at $t_1=\frac{2d_1}{c}$ where $c$ is speed of light. Similarly, the second impulse appearing at $t_2=\frac{2d_2}{c}$, corresponds to the light reflected from object $p_2$. The last impulse at $t_3=\frac{2d_3}{c}$ is due to the light traveled following 2 paths: ($C \rightarrow p_1 \rightarrow p_2 \rightarrow C$) and ($C \rightarrow p_2 \rightarrow p_1 \rightarrow C$). Clearly, these two paths have same travel distances, and thus in the time response they both appear at $t_3=\frac{d1+d2+d3}{c}$. By using these 3 values, $t_1, t_2, t_3$, one can find $d1, d2$ and $d3$. In other words, one can completely reconstruct relative positions of objects $p_1$ and $p_2$ up to Euclidean congruence.

Now, we fix the distance $d_3$ and keep increasing the distances $d_1$ and $d_2$. Naturally, at some point, the distance $d_3$ will be below the Rayleigh diffraction limit for the conventional imaging system, i.e. one can not visually separate two points ($p_1$ and $p_2$). However, for the proposed imaging system, the increase of the distances $d_1$ and $d_2$ results only in the time shift of the entire signal. The difference between 3 time tags, $t_1, t_2, t_3$, is conserved. Therefore, one can still recover the scene($d_1, d_2$ and $d_3$) up to Euclidean congruence.

Naturally, the arising question is whether it is possible to reconstruct the scene when the number of objects is larger. Is the reconstruction unique, i.e., is it possible that two or more configurations generate the same time response? Bellow, we show that the reconstruction of a point-cloud with $n$ objects is possible under some assumptions.

In this section, we use the results of Gkioulekas et al.~\cite{proof_pointcloud}, which shows that when the measured time-response of the scene contains a sufficiently rich set of first and second bounces we can reconstruct the point-cloud up to Euclidean congruence. Here we follow the same notation as~\cite{proof_pointcloud}. 

Now, consider a scenario where the scene consist of multiple point objects (Figure \ref{bounce_path}a). Let $ p_1, .. ,p_n \in \mathbb{R}^3 $ be $n$ point objects in the scene and $p_0\in \mathbb{R}^3$ be a position of our imaging system(laser and detectors are co-located). See Figure \ref{bounce_path}b. We call graph $ \textbf{S}= (p_1, .. ,p_n ) $ as a scene configuration and $ \textbf{K}= (p_0,p_1, .. ,p_n ) $ as a total configuration. The scene is illuminated with a pulsed laser and the detector observes returning signal from the scene. We define \textit{light path} $\alpha_k =[p_0,p_1,..,p_z,p_0](z \in \mathbb{N})$ as a finite sequence of points where light has traveled. Where $z$ is some integer. Note that any sequence $\alpha$ starts from $p_0$ and ends at $p_0$, since our light source and detectors are co-located. First bounce, \textit{ping}, is a light path $\alpha_k=[p_0, p_i, p_0]$ for $i=1,..,n$. Second bounce, \textit{loop}, is a light path $\alpha_k=[p_0, p_i, p_j, p_0]$ for $i \ne j$. Let $v_k$ be a length of $\alpha_k$. Then, our measurement contains ensemble $\beta=[v_1,v_2,...,v_k]$, where $v_k$ are returned first or second bounce.  

Next, let $\textbf{K}_5$ be a sub-graph of $\textbf{K}$ that has 5 vertices including $p_0$. If the measurement ensemble $\beta$ contains all pings and triangles of $\textbf{K}_5$ that starts and ends at $ p_0 $, then we say $\textbf{K}_5$ is contained within a $\beta$. See figure \ref{bounce_path}c. In this example, the measurement ensemble $\beta$ contains four pings ($\alpha=[p_0, p_i, p_0]$ for $i=1,..,4$) and six loops ($\alpha=[p_0, p_i, p_j, p_0]$ for $i,j \in (1,2,3,4)$ and $i \ne j$). Next, if the measurement ensemble $\beta$ contains a ping ($\alpha=[p_0, p_4, p_0]$) and three loops ($\alpha=[p_0, p_i, p_4, p_0]$ for $i=1,..,3$) then we say that $\beta$ allows for trilateration. See figure \ref{bounce_path}d. 

Finally, we can apply the theorem from Gkioulekas et al.~\cite{proof_pointcloud} which says that if $ \textbf{K}= (p_0,p_1, .. ,p_n ) $ is unknown configuration and $\beta$ is the measurement ensemble of $\textbf{K}$ that allows for trilateration, then one can find trilateration based process to reconstruct $ \textbf{K} $ up to Euclidean congruence. Readers can find detailed mathematical proof of the statement in ~\cite{proof_pointcloud}. Note that there exists a generic(or trivial) point configuration $\textbf{K}$ for which unique reconstruction is not possible. However, these special configurations occur very rarely~\cite{proof_pointcloud}.

\begin{algorithm}[!b]
\SetAlgoLined
\PrintSemicolon
\SetKwInOut{Input}{input}\SetKwInOut{Output}{output}
\SetKwFunction{CoreFinding}{CoreFinding}
\SetKwFunction{AddVertex}{AddVertex}
\Input{ Distance list $\beta$}
\Output{ Reconstructed point-cloud}
\BlankLine
\CoreFinding{}\\
\While{Core is not found}
{
Choose D=3 entries(random) from the distance list $\beta$, and test all possible two first and one second bounce pairs to construct base triangle \\
\If{a base triangle is found}
{
Choose D=6 entries from the distance list $\beta$, and test all possible two first and four second bounce pairs to construct two tetrahedra with the base triangle\\
\If{two tetrahedra are found}
{
choose D=1 entry from the distance list $\beta$, and do bridge bond test.\\
\If{bridge bond test is passed}{Core is found}
}
}
}
\BlankLine
\AddVertex{}\\
\While{Distance list $\beta$ is not empty}
{
Choose(random) four points(tetrahedra) from the substructure including the origin point\\
Choose D=3 entries from the distance list $\beta$, and test all possible one first and two second bounce pairs to construct tetrahedra with one side lie on the previously chosen tetrahedra\\
\If{tetrahedra is found}
{
Choose D=1 entry from the distance list $\beta$, and apply bridge bond test. \\
\If{bridge bond test is satisfied}
{Add tetrahedra to the existing substructure and remove used entries from $\beta$}
}
}

\Return point-cloud\;
\caption{Modified TRIBOND}
\end{algorithm}

\section{Reconstruction algorithm}
In this section, we design a trilateration based reconstruction algorithm for our imaging system. There exist algorithms(TRIBOND~\cite{UDGP} and LIGA~\cite{liga_Juhas}) that reconstruct a point-cloud given list of unassigned edge measurements. TRIBOND is a deterministic algorithm that addresses the unassigned distance geometry problem(uDGP). It has been successfully applied to reconstruct the structure of molecules and nanoparticles using edge distance lists extracted from X-ray or neutron diffraction data. However, we can not apply TRIBOND algorithm directly to our data, because it contains not only first bounce but also second bounce. Moreover these bounces are unlabeled. It requires modification. Gkioulekas et al.~\cite{proof_pointcloud} showed that it's possible to reconstruct point-cloud using unlabeled edge measurements. Here we modify the TRIBOND algorithm~\cite{UDGP} with method described in ~\cite{proof_pointcloud} to process data acquired by our imaging system. The modified TRIBOND algorithm consists of two parts: core finding and adding a vertex. 

\subsection{Core finding} 
First step of the buildup algorithm is to find the core. The core of the embedding point-cloud is a over-constrained set of 5 points including the source. See figure \ref{bounce_path}c. The core can be broken down into 3 pieces: the base triangle and two tetrahedra. The base triangle (Figure \ref{bounce_path}c $(p_0, p_1, p_3)$)  is constructed using two first (Figure \ref{bounce_path}c dashed green line) and one second(Figure \ref{bounce_path}a solid red line) bounces. Each tetrahedra (Figure \ref{bounce_path}a $(p_0, p_1,p_2, p_3)$ and $(p_0, p_1,p_4, p_3)$) uses one first and two second bounces. Since the bounces are not labeled, one has to exhaustively search over all possible first and second bounce pairs to build the base triangle and tetrahedra. Finally, we loop through all remaining bounces to find one second bounce to test and check if it fits into bridge bond $(p_2,p_4)$ between vertices's of two tetrahedra. If a correct second bounce is found and bridge bond is satisfied, we found a core structure and we can move to "adding a vertex". If the bridge bond is not satisfied, we restart "core finding" and choose another base triangle and tetrahedra by exhaustive search. This process is repeated until the core structure is found.

\subsection{Adding a vertex}
After core is found, next step is to iteratively add a vertex to the core. First, we choose a random tetrahedron from the current structure. The next step is to search over all possible combination of four distances from the remaining distance pool. Three distances will form one first and two second bounce distances. The remaining distance is used to test the bridge bond for the chosen rigid substructure. For instance, in figure \ref{bounce_path}d $(p_0, p_1,p_2, p_3)$ is chosen as a rigid substructure and $p_4$ is added point using one first bounce $[p_0,p_4,p_0]$ and two second bounces $[p_0,p_1,p_4,p_0]$ and $[p_0,p_2,0_4,p_0]$. The second bounce $[p_0,p_3,p_4]$ is used for the bridge bond check.

\iffalse
\begin{algorithm}
    \caption{Modified TRIBOND}
  \begin{algorithmic}[1]
    \REQUIRE Distance list $\beta$ is not generic
    \item[\textbf{Core finding:}]
    \WHILE{\textit{Core is not found}}
      \STATE Choose D=3 entries(random) from the distance list $\beta$, and test all possible two first and one second bounce pairs to construct base triangle
      \STATE if a base triangle is found, choose D=6 entries from the distance list $\beta$, and test all possible two first and four second bounce pairs to construct two tetrahedra with the base triangle
      \STATE if a base triangle and two tetrahedra are found, choose D=1 entry from the distance list $\beta$, and do bridge bond test.
      \STATE if bridge bond test is passed, then core is found
    \ENDWHILE
    
    \item[\textbf{Adding a vertex:}]
    \WHILE{\textit{Distance list $\beta$ is not empty}}
        \STATE Choose four random points(tetrahedra) from the substructure including the origin point
        \STATE Choose D=3 entries from the distance list $\beta$, and test all possible one first and two second bounce pairs to construct tetrahedra with one side lie on the previously chosen tetrahedra
        \STATE if a tetrahedra is found, choose D=1 entry from the distance list $\beta$, and apply bridge bond test. If bridge bond test is satisfied add tetrahedra to the existing substructure and remove used entries from $\beta$
    \ENDWHILE

  \end{algorithmic}
\end{algorithm}
\fi

\section{Simulations}
\begin{figure}
\centering
  \includegraphics[width=13cm, height=8cm]{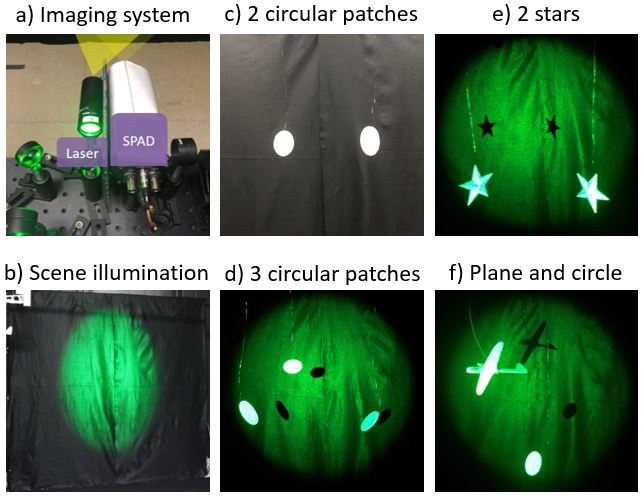}
  \caption{\textbf{Imaging system and examples of the scene}
  \label{fig_simulation_setup}
    \textbf{a,b)} The imaging system with co-located ultrafast pulsed laser and SPAD. We placed a diffuser in front of the laser to scatter the laser light into the scene. Note there is no lens before SPAD. On the right, the scene with illumination is shown. 
    \textbf{c,d,e,f)} Examples of the test scenes. All targets have a dimension approximately 4x4[cm], except the plane, which has dimensions 8x8[cm]. 
	}
\end{figure}

\begin{figure}
\centering
\begin{minipage}{.5\textwidth}
  \centering
  \includegraphics[width=1\linewidth]{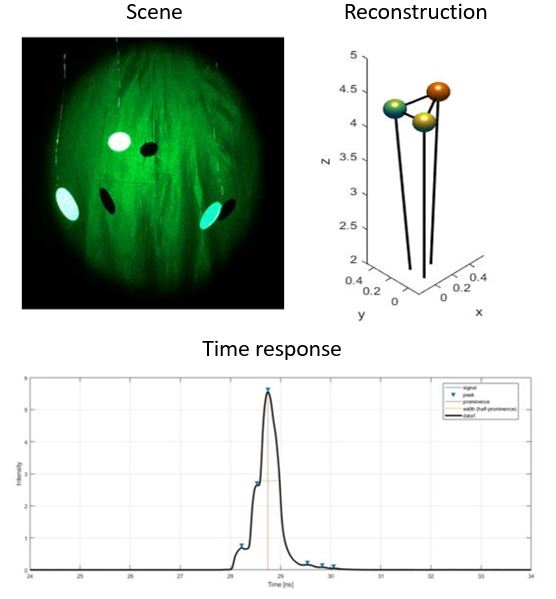}
  \captionof{figure}{\textbf{Real experiment with 3 white  patches.} 3 white patches with diameter 4[cm] are placed in the scene. In the time response, we can see 6 peaks in total: 3 first and 3 second bounces.   }
  \label{fig_3patches}
\end{minipage}%
\begin{minipage}{.5\textwidth}
  \centering
  \includegraphics[width=1\linewidth]{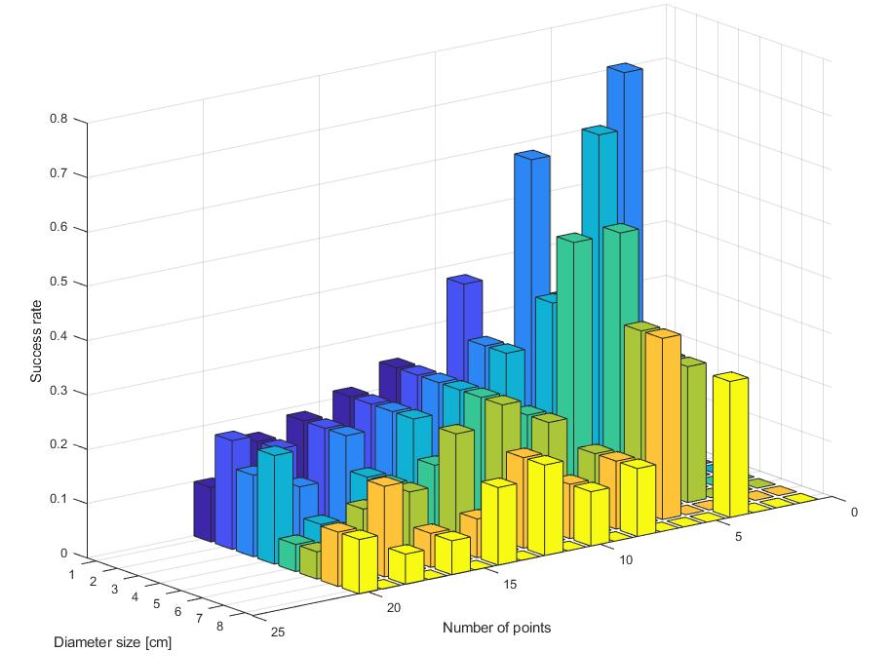}
  \captionof{figure}{\textbf{Resolution test using 2 circular patches} Statistical analysis of scene recoverability. The plot shows the probability of correctly imaging all targets in a scene of spheres randomly placed in a volume of 10 cubic meters. The number of spheres varies from 5 to 20 and the sphere diameter from 1 to 8 centimeters. The plot is generated from simulated data using the realistic parameters of detector and photon noise and assuming a 10 W illumination pulse from the laser.}
  \label{fig:test2}
\end{minipage}
\end{figure}

Here we test the modified TRIBOND algorithm using simulated data. We generate simulated data using a simplified version of the transient light transport renderer~\cite{zaragoza_render, hernandez2017computational}. The renderer is successfully used in many times of flight applications~\cite{zar_ex1,tsai2017geometry}. Similar to the previous chapter, we consider the following scenario. Let $ p_1, .. ,p_n \in \mathbb{R}^3 $ be n point objects in the scene and $p_0\in \mathbb{R}^3$ be a position of the TERA imaging system. Let $ d_i $ be the distance from $ p_i $ to $ p_0 $ for $ i=1..n $ and $ d_{ij} $ be the distance from $ p_i $ to $ p_j $ for all $ i \neq j $. See figure \ref{fig_simulation_setup}. Generated time response contains a mixture of first and second bounce.  Some of the signals can be missing because of occlusion or overlap. Figure \ref{fig_timeresponse_reconstruction} shows an example simulated time response. The scene contains $n=10$ point objects and located far away from the imaging system. The figure shows the entire time response and, separately, first and second bounces. Blue triangles represent locations of the peaks in the signal. We transform these locations of the peaks into a distance list and use it as an input to the modified TRIBOND algorithm. On the right side of Fig.\ref{fig_timeresponse_reconstruction} result of reconstruction is shown. The reconstruction matches the original point-cloud configuration up to Euclidean congruence. In other words, all pair-wise distances were recovered; however, the rotation and transpose of the point-cloud are still remaining unknown. Figure \ref{fig:test2} shows Statistical analysis of scene recoverability using multiple circular patches with different sizes.

\section{Proof of concept Experiments}
In this section, we demonstrate the performance of the algorithm using experimental data. The experimental setup is shown in the figure \ref{fig_simulation_setup}a. The imaging system contains an ultra-fast laser that emits laser pulses with 50[ps] width and 10MHz repetition rate. An example of the laser illumination area is shown in figure \ref{fig_simulation_setup}b. The wavelength of the laser is set to 532[nm]. The returning light is collected by a single-photon avalanche diode (SPAD), which has 20 micron diameter active-area. At wavelength 532[nm], it has a photon detection efficiency of 30$\%$. The total effective temporal impulse response of our imaging system is 80[ps]. Note that we don't use any lenses in front of the SPAD. 
\begin{figure}
    \centering
  \includegraphics[width=15cm]{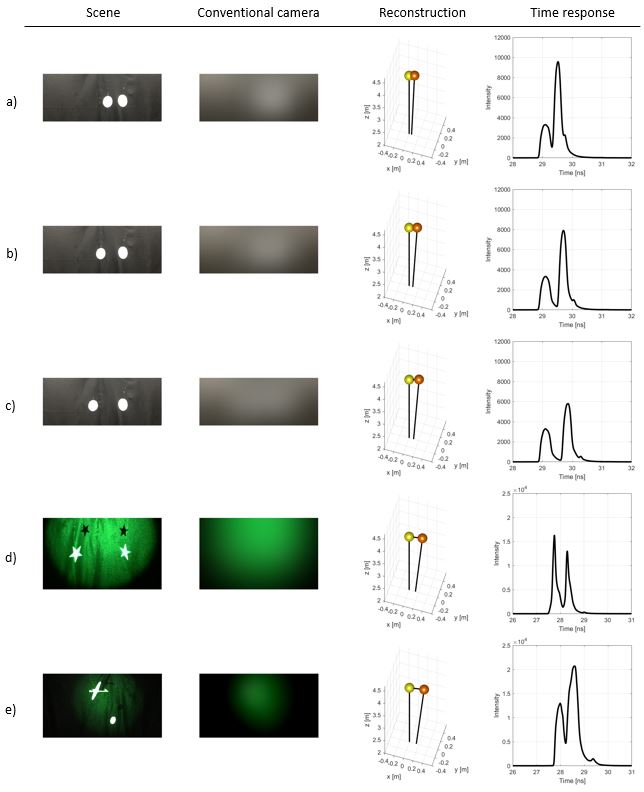}
  \caption{\textbf{Resolution test using 2 targets}
  \label{fig_all_recon}
  \textbf{a)} Circular patches with diameter 3[cm] are placed at distance 10[cm] from each other. In the time response, we can see clearly three peaks. Two first and one second bounce. Two first bounces have different intensity because of the angle orientation.   
  \textbf{b)} Circular patches with diameter 3[cm] are placed at distance 15[cm] from each other. 
  \textbf{c)} Circular patches with diameter 3[cm] are placed at distance 20[cm] from each other.
  \textbf{d)} Two star targets of size 4x4[cm] are placed at distance 12[cm] from each other. 
  \textbf{e)} Circular patch and a plane are placed at distance 15[cm] from each other.
	}
\end{figure}

The scene is located approximately 4[m] away from the system. Examples of test scenes are shown in the figure \ref{fig_simulation_setup}c,d,e,f.  Circular patches and starts have a size of 4x4[cm], and the plane is 8x8[cm]. First, we test the method using two circular patches. We place these two patches approximately 15 [cm] from each other and start moving apart up to 30[cm]. See figure \ref{fig_all_recon}a-c. The first column shows the actual scene picture. The second column is a simulated picture of the scene of using a camera that has an aperture size of 20 microns. The third column is a reconstruction of the scene using acquired data. Finally, fourth column shows the actual data. In the reconstruction, one can see that the two patches clearly are separated. Next, we test the method using targets(stars, plane, circular patch) with different shapes and materials(wooden star). See figure \ref{fig_all_recon}d,e. The shape of the peak is slightly different from the circular patches, but one can still find three peaks in the data. Lastly, in figure \ref{fig_3patches} reconstruction of 3 circular patches are shown. In the data, there are three first bounces, and three second bounces peaks.  

\iffalse
\begin{figure}[ht!]
  \centering
  \includegraphics[width=9cm]{Image/3patches.JPG}
  \caption{\textbf{Real experiment with 3 white  patches}
  \label{fig_3patches}
  %
  \textbf{} 3 white patches with diameter 4[cm] are placed in the scene. In the time response, we can see 6 peaks in total: 3 first and 3 second bounces.    
	% 
	}
\end{figure}

To estimate the average performance of the algorithm, we performed statistical analysis using simulation. The scene is placed at 100[m] away from the imaging system. We vary the number of small spherical objects in the scene from 5 to 20 and vary the diameter of the spheres from 1 to 10[cm], and randomly place them in the scene. Figure \ref{massive1} shows the results of the simulation. We can notice that as the number of objects and the size of the diameters increase the reconstruction success rate decreases due to the peaks overlap or missing.

\begin{figure}[ht!]
  \includegraphics[width=13cm,height=8cm]{Image/massive.JPG}
  \caption{\textbf{Resolution test using 2 circular patches}
  \label{massive1}
	%
    Statistical analysis of scene recoverability. The plot shows the probability of correctly imaging all targets in a scene of spheres randomly placed in a volume of 10 cubic meters. The number of spheres varies from 5 to 20 and the sphere diameter from 1 to 8 centimeters. The plot is generated from simulated data using the realistic parameters of detector and photon noise and assuming a 10 W illumination pulse from the laser.
	}
\end{figure}
\fi

\section{Conclusion}
In this paper, we present a novel imaging method that does not depend on the fundamental diffraction limit. The method makes use of hardware and algorithm to break the resolution limit under certain conditions of the scene (point-cloud assumption of the scene). In conventional imaging systems, the fundamental diffraction limit is governed by the size of the aperture, wave-length, and distance to the target. With fixed wave-length and distance, our proposed method does not depend on the size of the aperture, but instead, depends on the time resolution of the imaging system. The method is robust to change in surface materials. Currently, the main limitation is that the method is only applicable to the point-cloud scenes. Such sparse scenes commonly occur in aerial or space imaging scenarios. The paper gives base theory and introduces a method to use multiply scattered light (second bounce). As a next step, we are exploring possibilities of using multiply scattered light for continuous surfaces. As well as using machine learning to do the classification of objects below resolution limit using multiply scattered light.

\textbf{Funding.} This works was funded by Air Force Research Laboratory through program AFRL-AFOSR (FA9550-15-1-0208).

\textbf{Acknowledgments.} We appreciate help of Toan Le on the TERA hardware setup, and helpful discussion with Xiochun Liu and Atul Ingle.

\textbf{Conflicts of interest.} The authors declare no conflict of interest.

\bibliography{reflist}

\end{document}